\documentclass{emulateapj}
\usepackage{epsfig}
\usepackage{graphicx}
\shorttitle{Evidence of temporal variabilities from TCAF}
\shortauthors{Dutta \& Chakrabarti}

\begin{document}
\centerline{Accepted for publication in Astrophysical Journal; 23rd August 2016}
\title{Temporal variability from the two-component advective flow solution and its observational evidence}

\author{Broja G. Dutta\altaffilmark{1,3} and Sandip K. Chakrabarti\altaffilmark{2,3}}
%\affil{Department of Physics, Rishi Bankim Chandra College, Naihati, West Bengal, 743165, India}

\altaffiltext{1}{Department of Physics, Rishi Bankim Chandra College, Naihati, West Bengal, 743165, India}
\altaffiltext{2}{S. N. Bose National Centre For Basic Sciences, Block JD, Sector - III,
Salt Lake, Kolkata - 700098, India.}
\altaffiltext{3}{Indian Centre For Space Physics, 43 Chalantika, Garia Station Road, Kolkata - 700084, India.}

\begin{abstract}
In the propagating oscillatory shock model, the oscillation of the post-shock region, i.e., the Compton
cloud, causes the observed low-frequency quasi-periodic oscillations (QPOs). The evolution of QPO frequency is explained
by the systematic variation of the Compton cloud size, i.e., the steady radial movement of the shock front, which is
triggered by the cooling of the post-shock region. Thus, analysis of energy-dependent temporal properties
in different variability time scales can diagnose the dynamics and geometry of accretion flows around black
holes. We study these properties for the high inclination black
hole source XTE J1550-564 during its 1998 outburst and the low-inclination black hole source GX 339-4 during
its 2006-07 outburst using RXTE/PCA data, and we find that they can satisfactorily explain
the time lags associated with the QPOs from these systems. We find a smooth decrease of the time lag as a function of
time in the rising phase of both sources. In the declining phase the time lag
increases with time. We find a systematic evolution of QPO frequency and
hard lags in these outbursts. In XTE J1550-564, the lag changes from hard to
soft (i.e., from a positive to a negative value) at a crossing frequency ($\nu_C$) of $\sim 3.4$ Hz.
We present possible mechanisms to explain the lag behavior of high and low-inclination sources 
within the framework of a single two-component advective flow model (TCAF).
\end{abstract}


\section{Introduction}

The soft  X-ray transient (SXT) XTE J1550-564 is one of the most
interesting Galactic black hole candidates that has been studied over a broad range of wavelengths.
The typical outburst behavior of XTE J1550-564 during 1998 began and ended in a low/hard state similar to
the outburst profile of other black holes (e.g., GRO J1655-40, GX 339-4, etc.). This typical outburst behavior is
supposedly to be due to the sudden change in viscosity in the system (Hoshi 1979, Mandal \& Chakrabarti, 2010).
In Chakrabarti et al. (2009, 
hereafter, Paper I), a systematic study of the evolution of quasi-periodic oscillation (QPO) frequencies 
was carried out during XTE J1550's 1998 outburst, and it was shown that the variation of the QPO frequencies during
the rising and the declining phases could be understood by assuming that the centrifugal pressure supported 
shocks formed in the sub-Keplerian component moving in and out during the rising and declining
phases, respectively, and all the while oscillating at periods comparable to the infall time scale in the 
post-shock region (commonly known as CENBOL or CENtrifugal barrier dominated BOundary Layer). 
These oscillations are primarily due to resonances occurring between the cooling time scale
and the infall time scale in the CENBOL. Very recently, Chakrabarti et al. (2015) demonstrated that once 
the resonance sets in, it is likely to remain locked in resonance. 
Thus, the so-called `Propagatory Oscillatory Shock (POS)' model of the QPO evolution
may indeed be valid for all such outbursting candidates.  

Just as the QPO frequency variation gives us a clue to the geometry variation of the Comptonizing
electron cloud, the time or phase-lag variation should also have 
independent information not only about the variation of geometry, but also about the energy dependent 
physical processes which are responsible for the emission of photons from various regions of the
disc. In the present paper, we extend our study of XTE J1550-564 as presented in Papers I and II, to understand
the cause of peculiarities, if any, of the time-lag behavior and its energy dependence.

In X-ray binaries (XRBs), rapid variability in the X-ray emission on time-scales of 
milliseconds to seconds is a common and very complex phenomenon. 
It was proposed (Lightman \& Eardley 1974; Shakura \& Sunyaev, 1976) long ago that instabilities in
the standard accretion disk may cause the observed fast variability in X-ray binaries.
In order to match the observation, the time variation could exist over a wide range of 
time-scales that depend on the range of unstable radii (Lyubarskii, 1997).
Low-frequency quasi-periodic oscillations (LFQPOs) with frequencies ranging from a few mHz to 
$\sim 30$ Hz are observed in black hole X-ray binaries (BHBs) 
(see, van der Klis 2004; Remillard \& McClintock 2006). 
Another timing property is the manifestation of a time-lag which means a delay between
the Fourier components of the hard and soft light curves (i.e., time difference in arrival times
between hard and soft photons). Although time lags are observed in a large 
number of astronomical sources,
we did not find any satisfactory explanation in the literature as to the physics 
of the origin(s) of this phenomenon.

An early explanation of the time lag assumed that it was due to the Comptonization of soft 
seed photons by hot electrons, known as Compton reverberation (Payne 1980; Miyamoto 
et al. 1988), which naturally produces hard time lags. Several models have been proposed 
(e.g., Cui 1999; Poutanen 2001; Nowak et al. 1999) to explain the hard and soft lags  
observed in Galactic binary sources. The soft phase lags can be explained 
by a model (Bottcher \& Liang 1999; Lin et al.2000) in which  
the perturbation is assumed to propagate from the inner disk to the outer 
disk the soft phase lags can be explained. When perturbations propagate inward, 
a hard phase lag is produced. However, it is difficult to explain the transition 
frequency (e.g., 3.4 Hz for XTE J1550-564; 2.4 Hz for GRS 1915+105) 
where the lag converts from the hard type to the soft type.
Qu et al. (2010) found that the frequency and phase lag are both energy dependent  
for the  0.5-10  Hz  QPOs  in  GRS  1915+105. 
In XTE J1550-564, a similar type of complex variability is also observed (Cui et al. 2000).
Reig et al. (2000) proposed that  hard and soft lags are due to the mechanisms of Compton up-scattering and
down-scattering. Basically, these mechanisms depend on gradient of
temperature and optical depth of the plasma in the Comptonizing region.
For low optical depth, only the up-scattering, i.e., positive lags 
are possible. For a large optical depth diffusion plus down-scattering open up the possibility of negative lags. 
In the Compton up-scattering model (Lee et al. 2001) soft lag is explained due to the
Comptonization delays. Basically, these lags are due to the effects of the difference in 
the travel time of light and are comparable to the light crossing time of the objects.
This lag does not switch sign, however. 
Poutanen \& Fabian (1999) argue that the reflection of hard X-rays from the outer part 
of the accretion disk produces time delays that we may have already observed in Galactic black holes.
In this case, the disk should be flared and the break in the time lag Fourier spectra would
correspond to the size of the accretion disc. 

Recently, using XMM data, Uttley et al. (2011) found in GX 339-4 that the softer disc photon lead the power-law
photons at frequencies below $\sim 1$ Hz. The lag behavior switches at frequencies above $\sim 1$ Hz where
softer disc photons start to lag the harder power-law photons.
This behavior was interpreted as being due to the effect of the propagation (soft photon leads) to 
reverberation effects (soft photon lags). 
According to Arevalo \& Uttley (2006), at small disc radii, an instability in accretion rate   
will cause a small variation only in disk emission inside that radius. This disc emission variability is dominated by X-ray
heating effects which is due to the fluctuation in mass accretion at that radii.

In active Galactic nuclei, both hard and soft lags are explained with reverberation lag where reflection of Comptonized
photons from the disc (Zoghbi et al., 2010; Wilkins \& Fabian, 2013) are considered
Indeed, the switch in lag signature (a few tens milliseconds on short time scale variability) 
is consistent with the light-travel lags of reprocessed hard emission (power-law) 
from the soft (disc) emission at a few tens of 
gravitational radii. Since the Comptonization process could be the key reason for lags, it would be 
interesting to check if the size of this region, which is also responsible for QPOs, actually decides the 
amount of lags as the time lag is likely to be proportional to the size of the Compton cloud.

In the context of the 1998 outburst of XTE J1550-564, a study of the phase-lag evolution during the initial 
rising phase was carried out by Cui et al. (1999, 2000). The magnitude of the lag was 
found to increase with the total X-ray flux. It was found that the QPO becomes stronger 
at higher energies (Cui, 1999; Chakrabarti \& Manickam, 2000 for GRS 1915+105), 
which further supports the view that a QPO originates due to the oscillation of the Comptonizing region.  
although we consider only one each of high- and low-inclination sources, each with similar 
spectral and timing properties, the results may be general enough since there are 
many studies in the literature which indicate that
these properties do have some inclination dependence. Munoz-Darias et al. (2013) and 
Heil, Uttley \& Klein-Wolt (2015) showed that in higher inclination systems, there is
systematically harder X-ray power-law emission. The properties of low-frequency QPOs are also inclination dependent.
On the other hand, Heil et al. (2015) found that the amplitude of the broadband noise (subtracting the 
low-frequency QPOs) is no longer dependent on inclination, implying its correlation with the source structure of
the emitting regions.
Using Monte-Carlo simulations, Ghosh et al. (2011) also find that for the same disk flow properties that the 
spectrum changes with inclination angle.

There are yet other effects, such as the focusing of photons due to the gravitational bending
of light from regions close to the black hole horizon, which could also be important
and the delay introduced would be energy dependent as the energetic photons tend 
to be emitted from regions close to a black hole. In the present paper, for comparison, we study the 
time lag of one high inclination object namely, XTE J1550-564, and one low inclination object, namely, GX 339-4,
for comparison. As we have discussed above, the result is likely to represent the whole class of objects
with high and low inclinations, respectively. We explain the results of our observations by  
considering several effects on the time lag, namely, (i) repeated Compton scattering which introduces higher
lags for higher-energy photons and for Compton clouds of larger size before they escape,
(ii) reflection, and (iii) focusing due to gravitational bending.
The structure of our paper is the following. In the Section 2, we discuss the properties of the two 
black holes under consideration. In \S 3, we present observational data and our analysis procedure. Specifically,
we present the lag results for the XTE J1550-564 for which detailed studies of QPOs and spectral properties have
already been presented in Papers I and II. We also include a similar study for GX 339-4. In \S 4, we present 
discussions of our results and provide an explanation of the behavior of the time lag. 
Finally, in \S 5 we provide our concluding remarks.

\section{Introduction to the black hole candidates under consideration}

\subsection{XTE J1550-564}

The soft  X-ray transient (SXT) XTE J1550-564 is one of the most
interesting Galactic black hole candidate which has been studied over a broad range of wavelengths.
The typical outburst behavior of XTEJ1550-564 during 1998 starts and ends in a low/hard state which is similar to
the outburst profile of other black holes (e.g., GRO J1655-40, GX 339-4 etc.). This typical outburst behavior is
supposed to be due to the sudden change in viscosity in the system (Hoshi 1979, Mandal \& Chakrabarti, 2010).
The time lag varies during the initial onset phase of 1998 outburst (Cui et al. 1999, 2000). The magnitude
of lag increases with X-ray flux. The coherence is roughly constant and high (value is $\sim  1$).
Constant value is maintained up to the first harmonics of the Quasi Periodic Oscillations.

Properties of LFQPOs and
spectral variability during 1998 outburst were reported in Papers I \& II. 

\subsection{GX 339-4}

The black hole candidate GX 339-4 is a low mass X-ray binary having a primary of mass $\ge 6{\rm M}_{\odot}$
(Hynes et al. 2003; Munoz-Darias et al. 2008) and it was first detected by the MIT X-ray detector
on board OSO¿7 mission by Markert et al. (1973).
Since its discovery, the source has exhibited four outbursts at 2 to 3 years intervals.
The evolution of low-frequency quasi-periodic oscillations (QPOs) and  
spectral variability during 2010 outburst have been studied in detailed by
Debnath et al. (2010, 2015) with Two Component Advective Flow model (TCAF) proposed by 
Chakrabarti \& Titarchuk (1995; hereafter CT95) where they found that the 
QPO frequencies are monotonically increasing from 0.102 Hz to 5.69 Hz 
within a period of $\sim $26 days. They explained this evolution with the propagating oscillatory shock
(POS) solution and find the variation of the initial and final shock locations and strengths.
This behavior generally matches the values obtained from spectral fit with TCAF model (Debnath et al. 2015). 
With the TCAF model, a clear physical picture of what happens in an outburst emerges.

The only major difference between these two objects is the inclination angle. 
GX339-4 belongs to a class of low inclination ($< 60 \deg $) sources (Zdziarski et al. 1998) 
whereas XTE J1550-564 has a high inclination angle of $74.7 \deg \pm 3.8 \deg$ (Orosz et al. 2011).
However, both the sources exhibit a similar type of systematic evolutionary properties during
the outbursts. Here we use this rich data set to study the evolution of lag variability during the outbursts.

\section{OBSERVATIONAL ANALYSIS}

We analyzed {\it RXTE} public archival observations of BHB GX 339-4 during 2007 outburst 
and observations covering the rising and declining phases of QPOs in 1998 outburst 
of XTE J1550-564 (for details, see, Paper I), limiting
our analysis to observations when low frequency QPOs are observed. 
We produced background-corrected PCU2 rates in the Standard 2 channel bands
A = 4-44 (3.3-20.20 keV), B = 4-10 (3.3-6.1 keV) and C = 11-20 (6.1-10.2 keV)
and defined hardness ratio as C/B. We used Good Xenon, Event and 
Single Bit data modes which contain high time resolution data for timing 
analysis, which was performed using the GHATS software. For each
observation we produced PDS every 16s in the channel band 0-35 (2-15 keV). 
We averaged them to obtain an average PDS for each observation and we subtracted the 
Poissonian noise contribution (Zhang et al. 1995). The PDSs were normalized
and converted to fractional squared rms. The power spectra were then
fitted with a combinations of Lorenzian (see, Nowak 2000) using XSPEC v 12.0.

We also take the cross spectrum which is defined as, $C(j) = X_{1}(j)^\ast X_{2}(j)$,
where $X_{1}$ and  $X_{2}$ are the complex Fourier coefficients for
the two energy bands at a frequency $\nu_{j}$ and $ X_{1}(j)^\ast$
is the complex conjugate of $ X_{1}(j)$ (van der Klis et al. 1987).
The phase lag between the signals of two different energy bands at
Fourier frequency $\nu_{j}$ is, $\phi_j=arg\left[C(j)\right]$
(i.e.,  $\phi_j$ is the position angle of $C(j)$ in the complex plane.)
and the corresponding time lag is  ${\phi_j}/{2 \pi \nu}$.
An average cross vector $C$ is determined by averaging the complex values
for every stretches of time length. In our analysis,
we produced a phase lag spectrum for each observation in our sample,
dividing the data into two energy bands (2-5 keV \& 5-13 keV) and
extracting cross-spectra from 16s intervals, which are then averaged
yielding one phase-lag spectrum per observation.
Positive phase-lag indicates that the hard photons (5-13 keV) lag
the soft photons (2-5 keV). Following Reig et al. (2000), 
we calculate QPO phase lag as the average of the phase lags over
the interval  $\nu_c \pm FWHM$, where $\nu_c$ is the centroid frequency of
the QPO and FWHM is its full width at half maximum as measured through a
fit with a Lorenzian component.

\subsection{XTE J1550-564}
\begin{figure}
\centerline{\includegraphics[angle=00,scale=0.35]{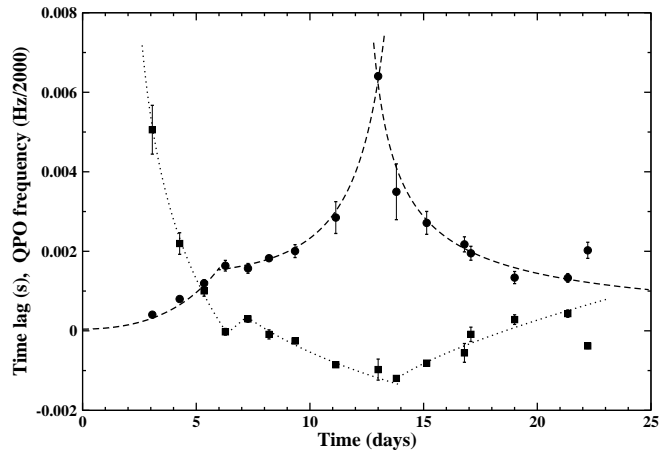}}
\caption{Evolution of time lag (filled squares) and QPO frequency (filled circles) 
are plotted with time (days) during rising (MJD 51065 to MJD 51076) and declining 
(MJD 51076 to MJD 51084) phases of XTE J1550-564 in the 1998 outburst. 
Time lag is calculated at QPO frequency ($f_Q$) integrated over the width which is equal
to FWHM ($F_w$) of the QPO itself.
}
\end{figure}
Figure 1 shows the systematic variation of the time lag (filled squares), which is QPO 
frequency-dependent and the QPO frequency (filled circles)
as a function of time (days) during the onset (MJD 51076-MJD 51084) and declining (MJD 51076-MJD 51084)
phases of the 1998 outburst in XTE J1550-564.
We consider the  $0^{th}$ day (i.e., MJD 51065) when the first QPO was detected. We studied the observations
that cover the first three weeks of the outburst. In the onset phase, we fit time lag variation with a curve 
where the time lag $\sim t^{-0.423 \pm .02}$ and the reduced ${\chi}^2$ is close to  $1.3\ ({\chi}^2/10)$. 
We find that time lag decreases with the increase of both the QPO frequency and time (day) 
during the onset phase. 
In the declining phase, we also find a systematic variation of the same time lag. 
The declining phase starts after observing the highest value of  $\nu_{QPO}=13.1$ Hz on  MJD 51076.
The fitted curve represents time lag $\sim t^{0.663 \pm .03}$ with a reduced ${\chi}^2$ close to $1.6\ ({\chi}^2/7)$.

\begin{figure}
\centerline{\includegraphics[angle=00,scale=0.35]{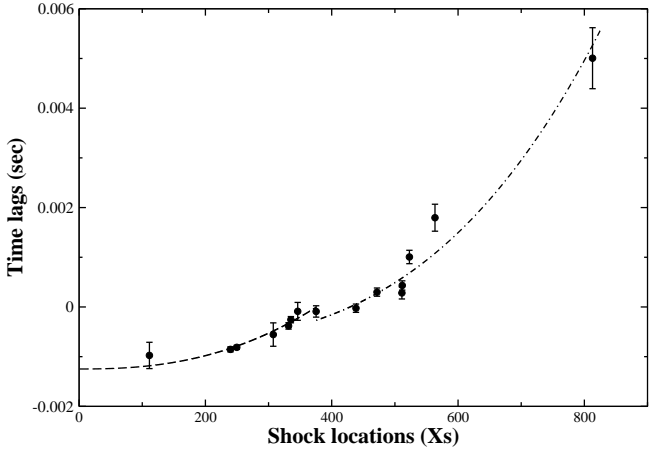}}
\caption{Variation of time lag with Shock location ($X_s$) is plotted
during the rising phase (MJD 51065 to MJD 51076) of XTE J1550-564 in 
the 1998 outburst. Time lag is calculated at the QPO frequency ($f_Q$)
integrated over the width equal to FWHM ($F_w$) of the QPO itself.}
\end{figure}

In the TCAF solution, 
the QPO frequencies are derived from the inverse of the infall time in the post-shock region. 
According to the POS solution (Chakrabarti et al. 2008, 2009;
Debnath et al. 2010, 2013), we can get an idea about the location of the shock wave ($X_s$)
from the observed QPO frequency ($\nu_{QPO}$) as it is believed that QPOs are generated due to the oscillations
of shock. The generated QPO frequency ($\nu_{QPO}$) is proportional to the inverse of infall time ($t_{infall}$, 
i.e., light crossing time from the shock location to the black-hole), i.e.,
$\nu_{QPO}  \sim (t_{infall})^{-1}$ and also $t_{infall} \sim  R X_s(X_s-1)^{1/2} \sim {X_s}^{3/2}$
$R$ is the shock strength ($={\rho_{+}}/{\rho_{-}}$, i.e., ratio of the post-shock to pre-shock densities.)
Thus, QPO frequency according to this model, $\nu_{QPO} \sim {X_s}^{-3/2}$. The time dependent
shock location is given by, ${X_s}(t) = r_{s0} \pm {vt}/{r_g}$, where $v$ is the velocity of the shock wave.
The positive sign in the second term is to be used for an outgoing
shock and the negative sign is to be used for the in-falling shock.  $X_s$ is the measured units
of the Schwarzschild radius  $r_g = 2GM/c^2$ where M is the BH mass and  $c$ is the velocity of light. 

Accordingly, the shock location is 
larger for lower QPO frequency. The opposite is also true. In Fig. 2, 
variation of time lag (in seconds) with Shock location ($X_s$) is plotted during 
the rising (MJD 51065 to MJD 51076) phases of XTE J1550-564 in the 1998 
outburst. We clearly see that purely from observational point of view
also, lag monotonically increases when QPO goes down and the derived shock location goes up. 
We thus have a fully consistent understanding that increase in 
QPO frequencies necessarily implies decrease in the size of the 
Comptonizing region. Even the `hiccup' on day 6 in QPO frequency shows up in the lag profile also. 

\begin{figure}
\centerline{\includegraphics[angle=00,scale=0.35]{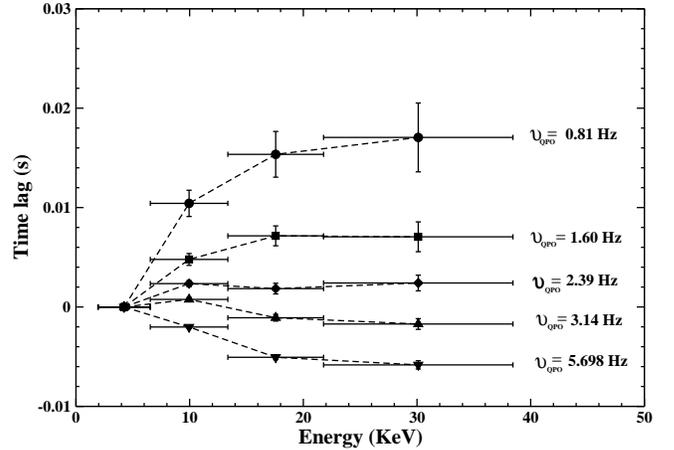}}
\caption{Energy dependent time lag (in seconds) for QPO centroid frequencies 
($f_Q =5.698$ Hz, $f_Q = 3.14$ Hz, $f_Q =2.39$ Hz, $f_Q =1.60$ Hz, $f_Q =0.81$ Hz) are plotted. 
We  averaged time lag over the width equal to FWHM ($F_w$) of the QPO itself. Here we calculated 
time lag w.r.t the reference energy band 1.94-6.54 keV (0-17, channels).
The other energy 
bands are 6.54-13.36 keV (18-36, channels), 13.36-21.78 keV (37-59, channels) and  21.78-38.44 keV (60-03, channels).}
\end{figure}

Because of the very nature of the physical processes by which high-energy photons are generated, 
the time lag must depend on the energy. First of all, repeated inverse Comptonization processes imply 
a higher time lag to generate higher energy photons. Similarly, focusing of emitted photons is also 
energy dependent since higher-energy 
photons are expected to come from regions closer to a black hole. This motivates up to compute energy
dependence of the lag. Figure 3 shows energy dependent time lag for QPO centroid frequencies $f_Q =5.698$ Hz,
$f_Q = 3.14$ Hz, $f_Q =2.39$ Hz, $f_Q =1.60$ Hz, $f_Q =0.81$ Hz. 
We  averaged time lag over the width equal to FWHM ($F_w$) of the QPO itself. Here we calculated
time lag w.r.t the reference energy band 1.94-6.54 keV (0-17, channels).
The other energy bands are $6.54-13.36$ keV (18-36, channels), $13.36-21.78$ keV (37-59, channels) 
and  $21.78-38.44$ keV (60-103, channels).
We find that lag could be positive or negative, depending on the QPO frequency and photon energy.
This behavior suggests that the contribution to lag from different mechanisms vary with 
shock locations. Time lag monotonically increases with energy for QPO frequency 0.81 Hz, but
monotonically decreases from a frequency of $\sim 3.4$ Hz onward.
The  complex  pattern  of  the  phase  lag  associated  with  the
QPO  bear remarkable resemblance to  that  observed  of  GRS 1915+105 
which is also high inclination source (Cui 1999, Reig et al. 2000).
We shall discuss the physical cause in the final Section.

\subsection{GX 339-4}
We now turn to a source which belongs to a low-inclination binary system. If focusing by 
gravitational photon bending is important in deciding the lag, low inclination sources are not
likely to be affected by this effect and the energy dependence of lag would look significantly 
different. This is precisely what we see in the source GX 339-4.

\begin{figure}
\centerline{\includegraphics[angle=00,scale=0.35]{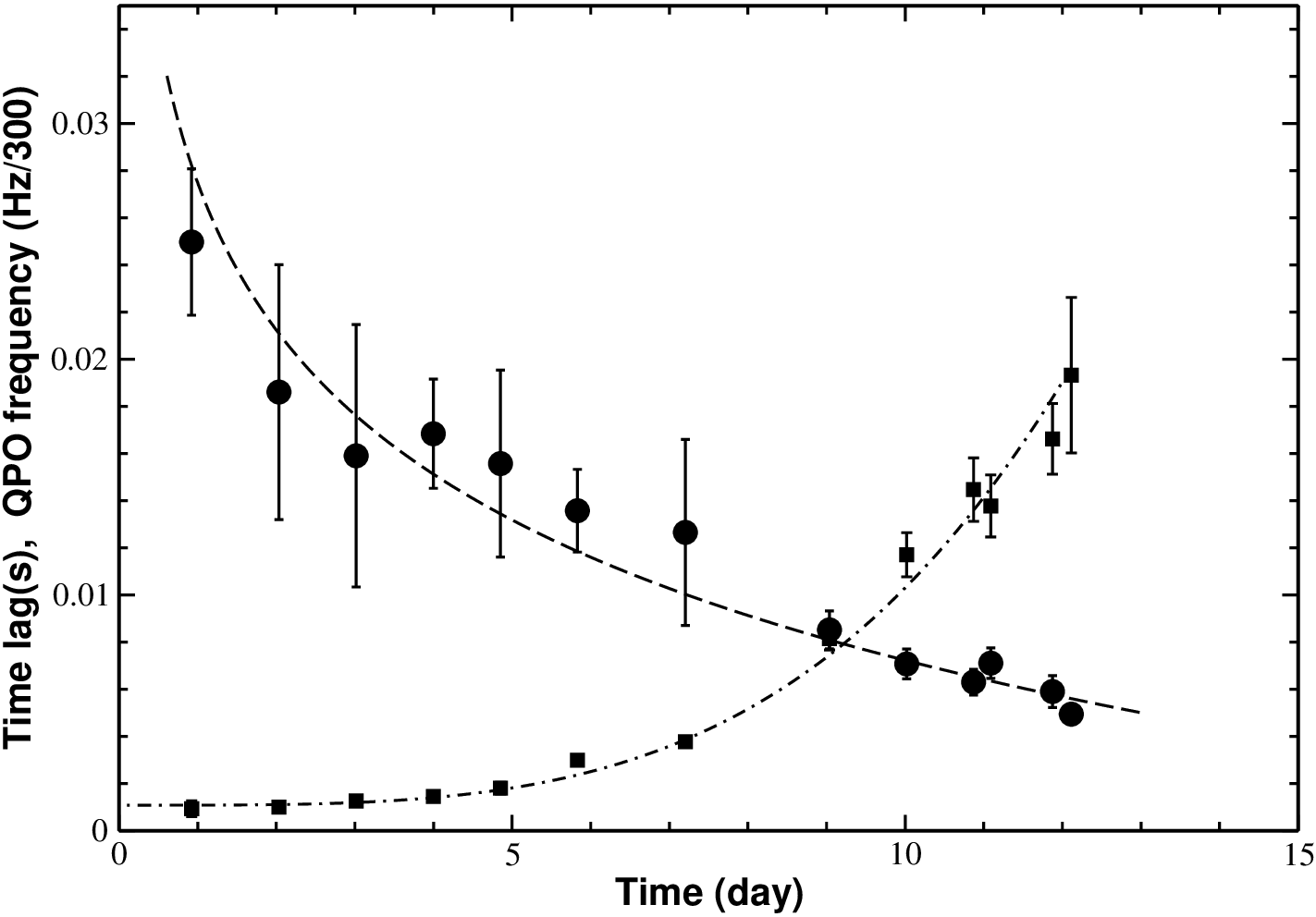}}
\caption{Evolution of time lag and QPO frequency are plotted as a function of time in days
during the rising (MJD 54133 to MJD 54145) phases of GX 339-4 in the 2007 outburst. 
Time lag is calculated at the QPO frequency ($f_Q$) by integrating over the range of the
FWHM ($F_w$) of the QPO.}
\end{figure}

Figure 4 shows a systematic variation of the frequency dependent time lag
as a function of day during the rising (MJD 51076 to MJD 51084) and the declining (MJD 54133 to MJD 54145) phases
of GX 339-4 during the 2007 outburst. Time lag is calculated at the QPO frequency ($f_Q$)
in the same way as before.

\begin{figure}
\centerline{\includegraphics[angle=00,scale=0.35]{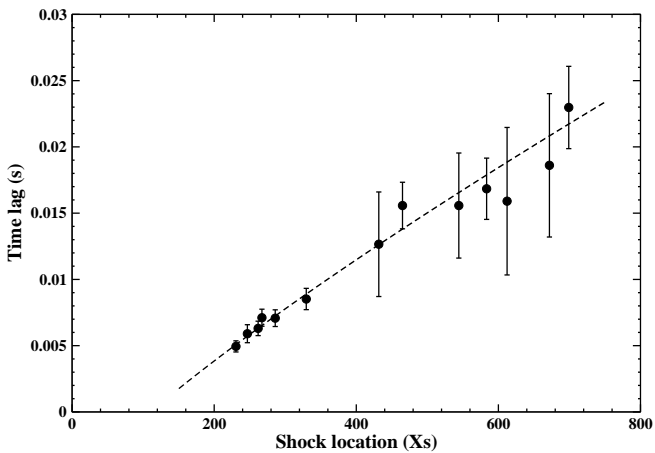}}
\caption{Variation of time lag with Shock location ($X_s$) 
during the rising (MJD 54133 to MJD 54145) phase
of GX 339-4 in the 2007 outburst. Time lag is calculated in the same way as in Fig. 2.}
\end{figure}

Figure 5 shows variation of time lag with the shock location ($X_s$) (computed from POS model) 
during the rising (MJD 54133 to MJD 54145) phase
of GX 339-4 during the 2007 outburst. The Figure suggests that as QPO frequency increases, location
of the shocks decreases, i.e., the size of the CENBOL is reduced systematically as the gradually weakening shock propagates
towards the black hole. 

Figure 6 shows energy dependent time lag for QPO centroid frequencies $f_Q =6.87$ Hz, $f_Q = 4.34$ Hz, $f_Q =4.14$ Hz,
$f_Q =1.297$ Hz, $f_Q =0.436$ Hz. The time lag is calculated as before. 
Here we choose 2.0-5.4 keV (0-12, channels) as a
reference energy band. The other energy bands are 5.4-6.9 keV (13-16, channels), 6.9-9.4 keV (17-22, channels) ,
9.4-13.1 keV (23-31, channels). We clearly find that energy dependent time lag depends on QPO 
frequency as well. We find hard lag to be monotonically increasing with energy of the photons for 
all the QPO frequencies. This behavior suggests that as the location of shock (i.e., QPO frequencies) 
changes, the contribution in total time lags that we observed from different mechanisms also 
changes. Furthermore, all the lags are positive, unlike the case of XTE J1550-564 which was of 
high inclination and contributions cancel to change sign at a particular shock location for a given object. 

\section{Discussion}

So far, we have seen that the time lag is not only a function of the average energy of the emitted 
photons, it is also a function of the inclination angle of the binary system. What is clear in both 
high and low inclination systems is that the lag increases when QPO frequency goes down, i.e., 
when the size of the Comptonizing region goes up. What is not obvious is the cause of 
change of sign at a specific QPO frequency, i.e., at a specific size. Below, we discuss the 
major processes which control the lag and give a possible reason of why the change of sign occurs 
only in high inclination systems.

\begin{figure}
\centerline{\includegraphics[angle=00,scale=0.35]{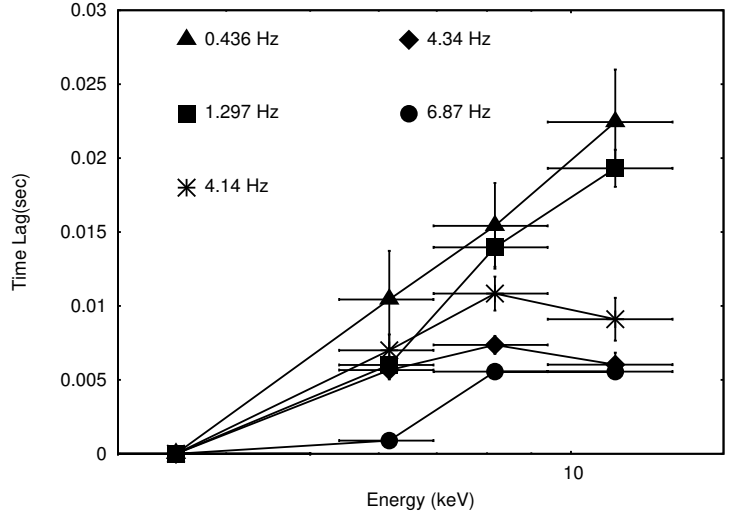}}
\caption{Energy dependence of time lag for different QPO centroid frequencies
($f_Q =6.87 Hz, f_Q = 4.34 Hz, f_Q =4.14 Hz, f_Q =1.297 Hz, f_Q =0.436 Hz$). 
Here we choose 2.0-5.4 keV (0-12, channels) as a reference energy band. The other
energy bands are 5.4-6.9 keV (13-16, channels), 6.9-9.4 keV (17-22, channels), 9.4-13.1 keV (23-31, channels).
Clearly varies on QPO frequency. We find that the hard lag is directly
related to the energy of photons for different QPO frequencies.}
\end{figure}

\subsection{Evolution of Times lag and Quasi Periodic Oscillations}

Typical evolutions of the QPOs in transient black-hole sources during outburst has already been established for
a long time (Debnath et al. 2008, 2010,2013; Dutta \& Chakrabarti 2010; Chakrabarti et al. 2005, 2008, 2009).
These evolutions suggest that certain specific physical mechanisms are in place which are responsible for 
the generation of QPOs for days after days (Chakrabarti et al. 2005, 2008, 2009;
Debnath et al. 2008, 2010, 2013). These authors found that the Comptonization 
region gradually shrinks in the rising phase of an outburst when QPO frequencies go up and 
the reverse is true in the declining phase. We would also expect that, in general,
there would be a monotonically decreasing lag with increasing QPO frequency and a monotonically increasing
lag with shock location. This is precisely what happens in both the objects: the net time lag 
due to Comptonization is the sum over light crossing times of the mean free paths $l_{Comp}$ 
between two successive Compton scatterings, i.e.,  $t_{lag}^c = \sum l_{Comp}/c  \propto X_s/c$.
However, when one considers the energy dependence, the picture is more complex. 
Since energy is expected to rise, after each successive scattering, higher energies are expected 
to have higher lags. As pointed out by earlier workers (Cui, 1999; Cui et al. 2000; Reig et al. 2000)
have pointed out, we also find that the energy dependence of the time lag is not straightforward 
to understand. In XTE J1550-564 (Cui et al.
2000) the lag first increases with frequency, peaks at some characteristic frequency, and then decreases 
and moves toward to negative (soft) lag at frequency $\sim 3.4$ Hz. This pattern of the phase lag 
associated with the QPO bears a remarkable resemblance to that observed in GRS 1915+105 
(Cui 1999; Reig et al. 2000), albeit with `zero-lag' occurring at a different QPO frequencies. 
This is due to the very nature of the emission process in both the Keplerian and sub-Keplerian components.

\begin{figure*}
\centerline{\includegraphics[angle=00,scale=0.50]{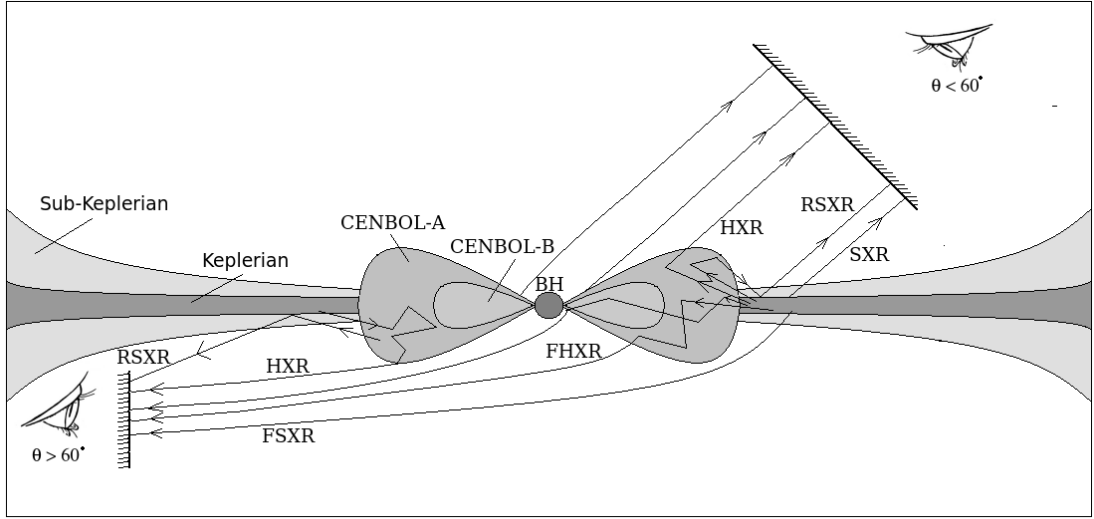}}
\caption{Cartoon diagram of two component advective flows where soft photons (SXR) from Keplerian discs 
are inverse Comptonized by the post-shock region, namely, Centrifugal pressure supported BOundary Layer (CENBOL)
and hard X-rays (HXR) are produced. Radiations could directly reach the observer, or reach through focusing
effects (FSXR, FHXR). Harder radiations from the inner regions of CENBOL (CENBOL-B), would be more focussed 
than the Comptonized photons which had undergone lesser number of scattering from outer CENBOL (CENBOL-A).
Reflected component (RSXR) from a Keplerian disc would primarily affect softer X-rays and may delay soft X-ray
travel time. Observers with $\theta \lesssim 60 \deg$ will see lesser focusing effects.
}
\end{figure*}

The physical processes contributing to lags may be understood from Fig. 7 where we show how the radiations 
from an accretion disc and Comptonizing cloud (CENBOL) may reach observers located at an angle 
$\theta \lesssim 60 \deg$ and $\theta \gtrsim 60 \deg$ through various paths. As already mentioned, 
the time of travel is proportional to the number of scattering at the CENBOL and one expects harder 
X-rays (HXR) to come at later times. However, radiations emitted 
from inner CENBOL (CENBOL-B) are also focussed due to gravitational bending of lights and may be further delayed.
Hard photons reflected from the Keplerian disc as softer radiation (RSXR) also takes a longer route, 
especially for high inclination sources. Thus, while each soft photon may have a range of time lag (high and low) 
depending on whether it is focussed and/or reflected before reaching the observer, hard X-rays are always expected 
to be delayed. This causes non-monotonicity of the lag with energy when the inclination angle is high.
Qualitative behavior could be understood from Fig. 8 where we schematically plot the time lag ($t_{lag}$) as a 
function of the size of the Comptonizing region ($R_c$), believed to be located at the inner region 
(Chakrabarti \& Titarchuk, 1995). We plot this for $\theta \lesssim 60 \deg$ on the left and for $\theta \gtrsim 60 \deg$ 
on the right. The boundary of this region (CENBOL) is the shock location $X_s$. For convenience, we assume
CENBOL is divided into two gross parts: CENBOL-A emits relatively softer photons than CENBOL-B. Lag due to 
Comptonization (superscript `c') of hard X-rays (HXR) would be proportional to the number of scattering which 
took place. Roughly, therefore, the energy would monotonically increase linearly with the size 
and its lag will also increase ($dt^c_{HXR}$ in Fig. 8). Soft X-rays from the Keplerian disc
will not be strongly affected by the size of the CENBOL. Hence, in Fig. 8 , $dt^K_{SXR} \sim 0$,
always. When $\theta \lesssim 60 \deg$, the focusing and reflection effects are also negligible.
So, we expect the lag $dt^c_{HXR} - dt^K_{SXR}$ to be always positive in this case (left panel of Fig. 8).
However, when the inclination is high ($\theta \gtrsim 60 \deg$), apart from the above two lags, we have 
monotonically decreasing effects of soft ($dt^F_{SXR}$, big circled curve; from CENBOL-A) and hard 
X-ray ($dt^F_{HXR}$ squared-curve; from CENBOL-B) lags due to focusing by gravitational bending effects.  
These are shown in the right panel of This is because bending is important only when the emission 
is closer to the black hole. Focusing is expected to cause a delay of the order of $1-2r_g/c$ for hard 
X-rays (from CENBOL-B) and $5-10r_g/c$ for soft X-rays (CENBOL-A) since they must be proportional to the 
size of the emitting region of corresponding radiations. Thus, the lag due to focusing of soft X-rays is 
on an average higher. The reflection component is a reprocessed CENBOL emission from the Keplerian disc and 
is a relatively softer radiation. The delay goes up linearly with the size 
of CENBOL ($dt^R_{SXR}$, small circled curve in right panel of Fig. 8) and the inclination angle. 
The combination of these four effects could result in interesting patterns. In the right panel of 
Fig. 8, we draw the total delay of hard X-rays (solid curve $dt^t_{HXR}$ which is the sum of $dt^c_{HXR}$ 
due to Comptonization and $dt^F_{HXR}$ due to focusing) and the total delay of soft X-rays 
(solid curve $dt^t_{SXR}$ is the sum of $dt^K_{SXR}$ from the Keplerian disc, $dt^c_{SXR}$ 
due to Comptonization, $dt^F_{SXR}$ due to focusing and $dt^R_{SXR}$ due to reflection), here we have 
assumed that no flare like situation (unless flashes like solar flares; rare in persistent sources) occurs where
hard X-rays could perhaps be expected earlier than the down scattered soft X-rays.
So we did not bring this issue at all. These two solid curves intercept 
at $R_c=R_{tr}$ below which the lag ($dt^c_{HXR} - dt_{SXR}$) is negative. This means that according to
TCAF, there is a cross-over QPO frequency (inverse of the infall time of matter from $R_{tr}$ to the horizon)
above which the hard lag would be negative. This cross-over frequency clearly depends on the mass of the black hole which determines
the length scales and time scales of the disc components and the CENBOL. 

\begin{figure*}
\centerline{\includegraphics[angle=00,scale=0.5]{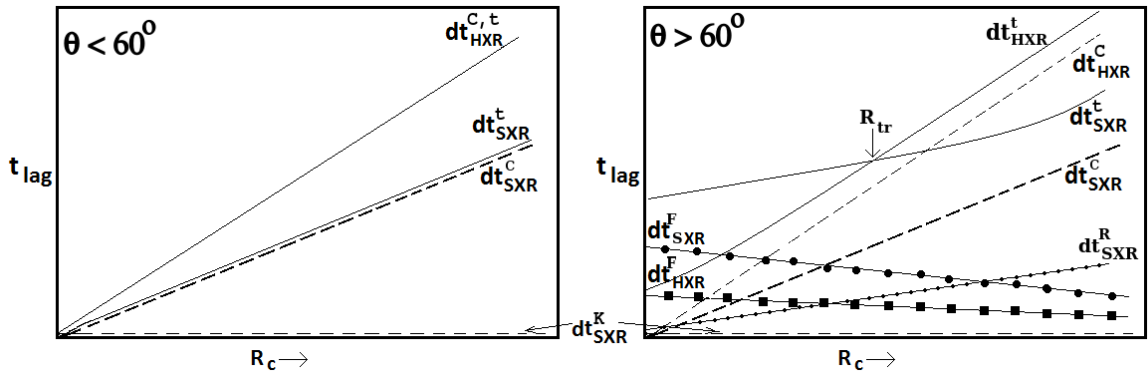}}
\caption{Schematic diagram explaining inclination angle and energy dependence of time lag. We consider
two cases: $\theta \lesssim 60\deg$ (left) and $\theta \gtrsim 60 \deg$ (right). In the former case, only Comptonization 
is important. In latter case, Comptonization, reflection and focussing are important. As a result, the
lag may change sign at a given frequency of the power density spectrum. See text for details.}
\end{figure*}

Heil et al. (2015) analyzed a large number of black hole candidates and
found that higher inclination sources with the same power spectral shape (in PDS) exhibit
systematically harder X-ray power-law emission. Though they found that the broad band noise in the power
density spectrum is independent of inclination angle, Type-C QPOs, located at different regions
of the spectrum are found to be inclination dependent. They conclude that this property strongly
depends on the inclination angle. In soft states, there are no low frequency QPOs and there is no
distinction with inclination angle. Munoz-Darias et al. (2013) found that marginally soft states will
be marginally harder in high inclination systems. In other words, hardening of spectrum is
due to higher abundance of harder radiation in relation to soft radiation. This is precisely what happens 
in Fig. 7. Ghosh et al. (2011) found that the same outgoing photons from Monte-Carlo simulations
when binned with respect to inclination angles, distinctly showed harder spectra at higher inclinations.

Our present work establishes this property more firmly using timing analysis. Here we 
considered two sources which show very similar systematic evolution in QPOs and spectral states.
In XTE J1550-564, QPO frequency rises from $0.01$ Hz to $14.0$ Hz and starts to fall in 
the declining state. During this time, the shock location also vary from $\sim 800$ 
to $\sim 100$. Also, from spectral studies with TCAF we find a similar changes in the
Compton cloud size (CENBOL). In GX 339-4, QPO frequencies vary from $0.10$ Hz to $5.69$ Hz and
then monotonically decreases, whereas the shock location changes from $\sim 600$ to $\sim 200$
(Debnath et al. 2010; Nandi et al. 2012). The complex outburst profiles of both the sources exhibit
similar spectral evolution: hard $\rightarrow$ hard-intermediate $\rightarrow$
soft-intermediate $\rightarrow$ soft $\rightarrow$ in rising state and in reverse
order in the declining state (Debnath et al. 2008). Moreover, they have comparable
masses while differing only in inclination which affected the QPOs and the phase
lags in the same way. Thus, our conclusion is expected to be valid for not just the two sources under consideration, 
but also for the two classes of sources, one with high and the other with low inclination.

\section{Conclusions}

In the present paper, we extended the earlier studies in Paper I and Paper II to understand the 
behavior of time/phase lag property of the black hole candidate XTE J1550-564.  Property 
of this high inclination object was contrasted to those of a low inclination black hole 
candidate, namely, GX 339-4. We concentrated on the lag at the QPO frequencies as the oscillation 
is more coherent and lag measurement is less erroneous. 
In this paper, we have used the TCAF paradigm for the explanation of time lag properties primarily because
only in this paradigm, there is a distinct prediction among time lag, size of the Compton cloud, QPO frequency
and spectral states. To our knowledge no other model in the literature such as the disk corona model (Haardt  \&  Maraschi,  1993;  
Zdziarski et al., 2003) and lamp-post models (Martocchia \& Matt   1996;   Miniutti \& Fabian   2004)
actually bring out such relations naturally. We found that in both the cases, the lag at QPO 
frequency generally rises as the frequency goes down. In fact,
exactly opposite result is found also in the declining phase. In a TCAF solution, this implies that the 
lag would increase as the size of the Comptonizing region increases. This is precisely what 
we see also. However, the dependence of lag on photon energy is more intriguing.
We find that GX 330-4 exhibits only positive lag for all QPO frequencies, while in XTE J1550-564, the lag switches sign
and it becomes negative above certain frequency. We discussed major effects which could be controlling the
property of the lag. Specifically, we showed that if we add up the qualitative variations of the lag components, 
then the high inclination objects could have negative time lags, i.e., soft photons appearing after hard 
photons due to reflection and focussing effects. We see this effect in XTE J1550-564 
at frequencies higher than $\sim 3.4$Hz. This frequency gives rise to a characteristic length-scale ($R_{tr}$)
where the lag changes its sign. Most certainly, this frequency is not universal 
as the cancellation of lags would depend on inclination angles and the mass of the 
black holes which determine the length scales. Indeed,
in another object, namely, GRS1915+105, having a mass at least twice as much, the transition frequency happens to be 
lower ($2.2$ Hz). Detailed work is in progress and would be presented elsewhere.

\section{Acknowledgements}
This research has made use of RXTE data obtained through HEASARC Online
Service, provided by the NASA/GSFC, in support of NASA High Energy Astrophysics Programs. 
We thank T. Belloni for providing the timing analysis software GHATS at X-ray Astronomy School (IUCAA, Pune). Also,  
B. G. Dutta acknowledges the support of UGC (Minor Research Project) to carry forward his research works.


\begin{thebibliography}{}
%\def\ref#1\par{\parshape=2 0in 14.5cm 1cm 13.5cm {#1} \par}
%\parskip=0pt
%\parindent=0pt
\bibitem{}Arevalo P. \& Uttley P., 2006, MNRAS, 367, 801
\bibitem{}Bottcher, M. \& Liang, E. P., 1999, ApJL, 511, L37
\bibitem{}Chakrabarti, S. K., Nandi, A., \& Debnath, D., et al., 2005, IJP, 79(8), 841 (arXiv:astro-ph/0508024)
\bibitem{}Chakrabarti, S.K., Debnath, D., \& Nandi, A., et al., 2008, A\&A, 489, L41
\bibitem{}Chakrabarti, S.K., Dutta, B.G. \& Pal, P.S., 2009, MNRAS, 394, 1463C
\bibitem{}Chakrabarti, S.K. \& Titarchuk, L.G., 1995, ApJ, 455, 623
\bibitem{}Chakrabarti, S.K., Mondal, S., Debnath, D., 2015, MNRAS, 452, 3451C
\bibitem{}Chakrabarti, S.K. and Manickam, S. 2000, ApJ, 531, 41
\bibitem{}Cui, W., Chen, W., \& Zhang, S. N., 1999, ApJ, 484, 383
\bibitem{}Cui, W., 1999, ApJ, 524, L59
\bibitem{}Cui, W., Zhang, S. N., \& Chen, W., 2000, ApJ, 531, L45
\bibitem{}Debnath, D., Chakrabarti, S. K. \& Nandi, A., 2010, A\&A, 520A, 98D
\bibitem{}Debnath, D., Chakrabarti, S. K., \& Nandi, A., et al., 2008, BASI, 36, 151
\bibitem{}Debnath, D., Chakrabarti, S.K., \& Nandi, A., 2013, AdSpR, 52, 2143
\bibitem{}Debnath, D., Mondal, S., \& Chakrabarti, S.K., 2015, MNRAS, 447, 1984
\bibitem{}Dutta, B.G. \& Chakrabarti, S. K., 2010, MNRAS, 404, 2136D
\bibitem{}Ghosh, H., Garain, S. K., Giri, K., \& Chakrabarti, S. K., 2011, MNRAS, 416, 959G
\bibitem{}Haardt, F., \& Maraschi, L., 1993, ApJ, 413, 507
\bibitem{}Heil, L. M., Uttley, P. \& Klein-Wolt, M., 2015, MNRAS, 448, 3348
\bibitem{}Hoshi, R., 1979, Progress of Theoretical Physic, 61,  1307
\bibitem{}Hynes R. I., Steeghs D., Casares J., Charles P. A., OBrien K., 2003, ApJ, 583, L95
\bibitem{}Lee, H. C., Misra, R.\& Taam, R. E., 2001, ApJ, 549L, 229L
\bibitem{}Lightman, A. P.\& Eardley, D. M., 1974, ApJ, 187, L1
\bibitem{}Lin, D., Smith, I. A., Liang, E. P. \& Böttcher, M., 2000, ApJ, 543L, 141
\bibitem{}Lyubarskii, Yu. E., 1997, MNRAS, 292, 679
\bibitem{}Mandal, S. \& Chakrabarti, S.K., 2010, ApJ, 710, L147
\bibitem{}Markert T. H., Canizares C. R., Clark G. W., Lewin W. H. G., Schnopper H. W. et al., 1973, ApJ, 184, L6
\bibitem{}Martocchia, A. \& Matt, G. 1996, MNRAS, 282, L53
\bibitem{}McClintock, J. E., \& Remillard, R. A., 2006, in Compact Stellar X-ray
Sources, ed. W. Lewin \& M. van der Klis, 39, 157
\bibitem{}Miniutti, G. \& Fabian, A. C. 2004, MNRAS, 349, 1435
\bibitem{}Miyamoto S., Kitamoto S., Mitsuda K., Dotani T., 1988, Nat, 336, 450
\bibitem{}Muñoz-Darias, T., Casares, J. \& Martínez-Pais, I. G., 2008, MNRAS, 385, 2205
\bibitem{}Munoz-Darias T., Motta S., Belloni T. M., 2011, MNRAS, 410, 679
\bibitem{}Muñoz-Darias, T., Coriat, M. \& Plant, D. S., 2013, MNRAS, 432, 1330
\bibitem{}Nandi, A., Debnath, D., \& Mandal, S., et al., 2012, A\&A, 542, 56
\bibitem{}Nowak, M. A., Wilms, J., Dove, J, B. 1999, ApJ, 517, 355
\bibitem{}Nowak, M. A., 2000, MNRAS, 318, 361
\bibitem{}Orosz, J. A., Steiner, J. F., McClintock, J. E., Torres, M. A. P., \& Remillard, R. A., et al., 2011, ApJ, 730, 75O
\bibitem{}Payne, D. G., 1980, ApJ, 237, 951
\bibitem{}Poutanen, J., 2001, in X-RAY ASTRONOMY: Stellar Endpoints,AGN, and the Diffuse X-ray Background. Edited by Nicholas E. White, Giuseppe Malaguti, and Giorgio G.C. Palumbo. AIPC, 599, 310
\bibitem{}Poutanen, J. \& Fabian AC, 1999, MNRAS, 306, L31
\bibitem{}Qu, J. L., Lu, F. J., \& Lu, Y. et al., 2010, ApJ, 710, 836
\bibitem{}Reig, P., Belloni, T., van der Klis, M., Mendez, M. \& Kylafis, N. D., 2000, ApJ, 541, 883
\bibitem{}Remillard R., McClintock J., 2006, ARA\&A, 44, 49
\bibitem{}Shakura, N. I.\& Sunyaev, R. A., 1976, MNRAS, 175, 613
\bibitem{}Uttley, P., Wilkinson, T. \& Cassatella, P. et al. 2011, MNRAS, 414L, 60
\bibitem{}van der Klis, M., Hasinger, G., Stella, L., Langmeier, A., van Paradijs, J., \& Lewin, W. H. G. 1987, ApJ, 319, L13
\bibitem{}van der Klis M., 2004, Advances Space Res., 34, 2646
\bibitem{}Wilkins, D. R. \& Fabian, A. C., 2013, MNRAS, 430, 247
\bibitem{}Zdziarski, A, A., Poutanen, J., Mikolajewska, J., Gierlinski, M., \& Ebisawa, K., et al.,1998, MNRAS, 301, 435
\bibitem{}Zdziarski, A. A., Lubinski, P., \& Gilfanov, M., et al., 2003, MNRAS, 342,355
\bibitem{}Zhang, W., Jahoda, K., Swank, J. H., Morgan, E. H. \& Giles, A. B., 1995, ApJ, 449, 930
\bibitem{}Zoghbi, A., Fabian, A. C., Uttley, P., Miniutti, G.,\& Gallo, L. C et al., 2010, MNRAS, 401, 2419
%%
\end{thebibliography}
\end{document}